\UseRawInputEncoding
\documentclass[referee]{raa}            % referee version: for submission

\usepackage{graphicx,times}             %for PS/EPS graphics inclusion, new
\usepackage{natbib}
\usepackage{amssymb,amsmath}
\usepackage{multirow}
\usepackage{booktabs}

\bibpunct{(}{)}{;}{a}{}{,}

\usepackage[a4paper=true,dvipdfm=true,pagebackref=true]{hyperref}
\hypersetup{colorlinks = true, linkcolor = green, anchorcolor = red, citecolor = blue, filecolor = red, pagecolor = red, urlcolor = red}

\graphicspath{{figures/}}
\newcommand{\tabincell}[2]{\begin{tabular}{@{}#1@{}}#2\end{tabular}}

\begin{document}

   \title{Ground Experiments and Performance Evaluation of the Low-Frequency Radio Spectrometer Onboard the Lander of Chang'e-4 Mission
%\footnotetext{$*$ Supported by the National Natural Science Foundation of China.}
}
%   \subtitle{I. Place Your Subtitle Here}

   \volnopage{Vol.0 (20xx) No.0, 000--000}      %%preserved for Editor. DOn't remove!
   \setcounter{page}{1}          %%starting page, preserved for Editor. DOn't remove!

   \author{Xin-Ying Zhu
      \inst{1}
   \and Yan Su
      \inst{1,3}
   \and Yi-Cai Ji
      \inst{2,3}
   \and Hong-Bo Zhang
      \inst{1,3}
   \and Bo Zhao
      \inst{2}
   \and Jun-duo Li
      \inst{1}
   \and Shun Dai
      \inst{1}
   \and Xi-Ping Xue
      \inst{1}
   \and Chun-Lai Li
      \inst{1,3}
   }

   \institute{Key Laboratory of Lunar and Deep Space Exploration, National Astronomical Observatories,
Chinese Academy of Sciences, Beijing 100110, China;  {\it suyan@bao.ac.cn}\\
        \and
              Institute of Electronics, Chinese Academy of Sciences, Beijing 100190, China;{\it ycji@mail.ie.ac.cn}\\
        \and
              University of Chinese Academy of Sciences, Beijing 100049, China;\\
       \vs\no
   {\small Received~~20xx month day; accepted~~20xx~~month day}}

\abstract{ The Low-Frequency Radio Spectrometer (LFRS) is a scientific payload onboard the Chang'e-4 lunar lander launched in December 2018. The LFRS provides in-situ measurements of the low-frequency radio phenomena on the far-side of the Moon for the first time in human history. To evaluate the performance of the LFRS, a series of ground experiments are conducted using a engineering model of the LFRS. It is not easy to perform the experiments because the EMI\footnote{ Abbreviation: EMI, Electro Magnetic Interference} from the Chang'e-4 lunar lander itself and the environment is very intense. The results after EMI mitigation show that the sensitivity of the LFRS may be $10\mathop{{}}\nolimits^{{-18}}Wm\mathop{{}}\nolimits^{{-2}}Hz\mathop{{}}\nolimits^{{-1}}$.
\keywords{Moon:far-side---space vehicles: instruments---techniques: spectroscopic}
}

 %%  \authorrunning{X.-Y. Zhu et al.}            %author_head in even pages
   \titlerunning{Ground Experiments and Performance Evaluation of the LFRS  }  % title_head in odd pages

   \maketitle
%% The author head (on even pages) and the title head (on odd pages) will be
%% automatically extracted from \author{} and \title{}. Whenever the title is too long,
%% you will be asked to supply a shorter one by inserting either \authorrunning{} or
%% \titlerunning{} before \maketitle. Anyway, you can specify your own heads.
%%
%%
%% Note: In the following text body of your manuscript, please note several differences from
%%       other major journals:
%% (1) \subsection{Please Capitalize the First Letter of Each Notional Word in Subsection Title}
%% (2) Please Capitalize the First Letter of Each Notional Word in all tables' captions

%
%________________________________________________ sections below
%
\section{Introduction}           %% first-level sections will be auto-capitalized
\label{sect:intro}

The atmosphere is not perfectly transparent at any radio frequency. Radio waves are scattered and absorbed by Earth's atmosphere. Low-frequency astronomical observation from ground is limited by severe ionospheric distortions below 50MHz and complete reflection of radio waves below 10-30MHz (\cite{Jester+2009}). Even from an orbiter around Earth, human-made interference from the Earth and natural radio emission from the Sun turn out to be too overwhelming for any observations in this frequency range. For these reasons, the low-frequency end is indeed one of the last portions of the electromagnetic spectrum to remain
terra incognita in astrophysics.(\cite{Boonstra+2010,Bentum+2011,Jester+2009,Takahashi+2003,Wolt+2012}).

Chang'e-4 probe is the first mission landed on the far-side of the Moon in human history. The probe was launched at 18:23 (UTC) on December 7th 2018, and landed successfully in the Von K¨¢rm¨¢n crater within the South Pole-Aitken (SPA) basin at 2:26 (UTC) on January 3rd 2019. It will attempt to collect new evidence from the most massive crater in the solar system to determine the age and composition of an unexplored region of the Moon, as well as develop technologies required for the later stages of the program(\cite{Li+2019,Wu+2017,Wu+2019}).
The LFRS is a scientific payload onboard the Chang'e-4 lunar lander. The primary motivation for the LFRS is to learn about the universe through low-frequency spectral window. The Moon can be utilized as a shield against unwanted radiations from the Earth. By taking advantage of the unique environment, many astrophysical topics of interest such as cosmology with HI line emission, solar and planetary radio bursts, local plasma environment above the Moon¡¯s
surface, ultra-high energy particle detection,meteoritic impacts could be studied through low-frequency observations(\cite{Jester+2009,Lazio+2011,Wolt+2012}). According to the characteristics of the Chang'e-4 mission, the main scientific objectives of the LFRS are to probe solar radio bursts and local plasma environment above the Moon¡¯s
surface in the frequency range 0.1-40MHz.

The Chang'e-4 probe was initially built as a backup for Chang¡¯e-3 and became operational after Chang'e-3 successfully landing in 2013. According to EMC\footnote{ Abbreviation:EMC, Electro Magnetic Compatibility} test results of Chang'e-4, the EMI from the Chang'e-4 lunar lander itself is very intense, so all target radio emissions are hidden in the EMI noise. A method is proposed in order to suppress the significant interference from the lander. Experiments on the ground (On Earth) are an essential tool for evaluating the payload performance, so a series of ground experiments were performed for this reason. In addition, the EMI mitigation method was also verified during the ground experiments.

%% Authors can give a citation as 'Michel et al. 1992'.
%% You may also use \cite, \citep and \citet for citation, and use Table~1 or Figure~1
%% and so forth. Using \ref and \label for cross-references of Tables/Figures
%% is a good way in adjusting/adding/removing text, tables or figures.

\section{Instrument Description}
\label{sect:Des}

\begin{figure}
   \centering
   \includegraphics[width=0.5\textwidth, angle=270]{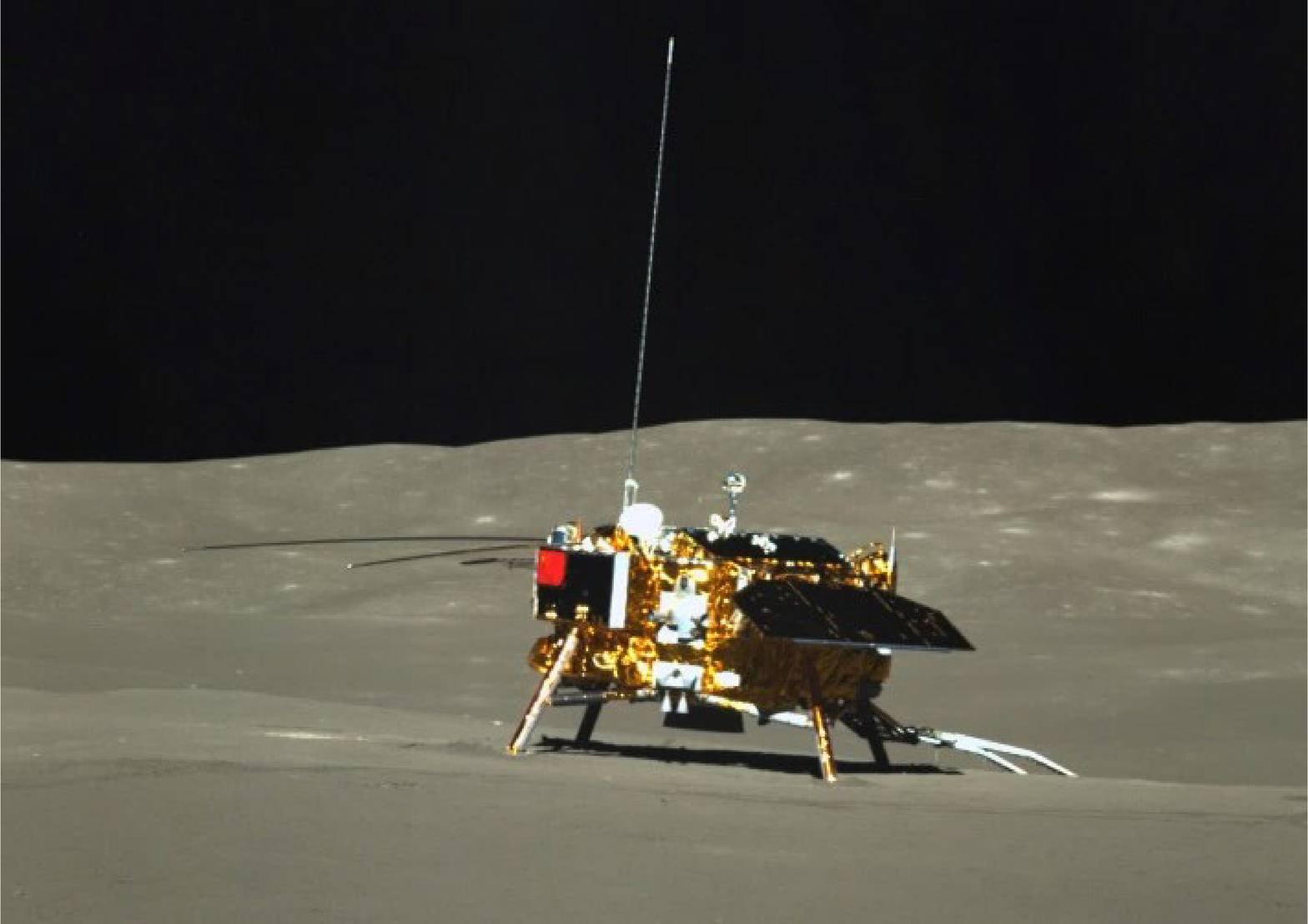}
   \caption{{\small The actual working environment of the LFRS on the Moon. This photo was taken by the Chang'e-4 cruiser.} }
   \label{fig:ms2019-0316fig1}
  \end{figure}

The LFRS which is mounted on the top of the Chang'e-4 Lander, was designed by IECAS\footnote{ Abbreviation:IECAS, Institute of Electronics, Chinese Academy of Sciences} together with NAOC\footnote{ Abbreviation:NAOC, National Astronomical Observatories, Chinese Academy of Sciences}, and manufactured by IECAS. The actual working environment of the LFRS on the Moon is showed in  \autoref{fig:ms2019-0316fig1}. This picture was taken by the Chang'e-4 cruiser.

The LFRS consists of sensors, pre-amplifiers, electronics unit and cable assemblies. The sensors and pre-amplifiers are installed outside the lander cabin, while the electronics unit is installed inside. The sensors are three 5m long antennas which are mounted on the top of the lander orthogonally(\cite{Arts+2010}). Another 20cm long antenna is mounted near the root of the long antenna as an auxiliary reference antenna. The signals of the above four antennas are pre-amplified and fed to the electronics unit. The electronics unit includes a control unit, a four-channel radio receiver covering the full band of 0.1$-$40MHz, and a calibration unit etc(\cite{Ji+2017}).

The three 5 meter long antennas receive both the scientific radio signals as well as the EMI of the lander, while the 20 cm long antennas only receive the EMI of the lander. The data of the 20 cm long antennas will be used to suppress the EMI in the off-line data processing stage. The specifications of the LFRS  are summarized in \autoref{Tab:Prin-LFRS}.
% reversion by 2020.3.5

%
%               one-column-spanning table
%________________________________________ Table 2: Use_of_the routines
\begin{table}
\begin{center}
\caption[]{{\small  Specifications of the LFRS instrument.}}\label{Tab:Prin-LFRS}

%%Please Capitalize the First Letter of Each Notional Word in table's caption

 \begin{tabular}{clcl}
  \hline\noalign{\smallskip}
No &  Parameter      & Characteristics                    \\
  \hline\noalign{\smallskip}
  1	 & Frequency            & $\mbox{0.1-40MHz}$              \\
  2	 & Receiver sensitivity & $\leq\mbox{10nV}/\sqrt{\mbox{Hz}}$               \\
  3	 & Dynamic Range        & $\geq\mbox{75dB}$                    \\
  4	 & Frequency Resolution & \tabincell{c}{$\leq\mbox{10kHz} \left(\mbox{0.1-2.0MHz}\right)$ \\$\leq \mbox{200kHz } \left(\mbox{1.0-40MHz}\right)$ } \\
  5	 & Max allowed bit rate         & $\mbox{5Mbps}$\\
  6	 &   Power              &  $   \leq\mbox{24W}$             \\

  \noalign{\smallskip}\hline
\end{tabular}
\end{center}
\end{table}

\section{Ground Experiment}
\label{sect:exp}
The main purposes of the ground experiments were to evaluate the performance of the prototype model of the LFRS and verify the off-line data processing method. According to the EMC test results of the Chang'e-4 lander, the detection capability of the LFRS is  related to its technical performance and depends on the EMI mitigation method for noise suppression of the lander. Therefore, the EMI mitigation method is also tested in the ground experiments.

  \begin{figure}
   \centering
   \includegraphics[width=0.5\textwidth, angle=270]{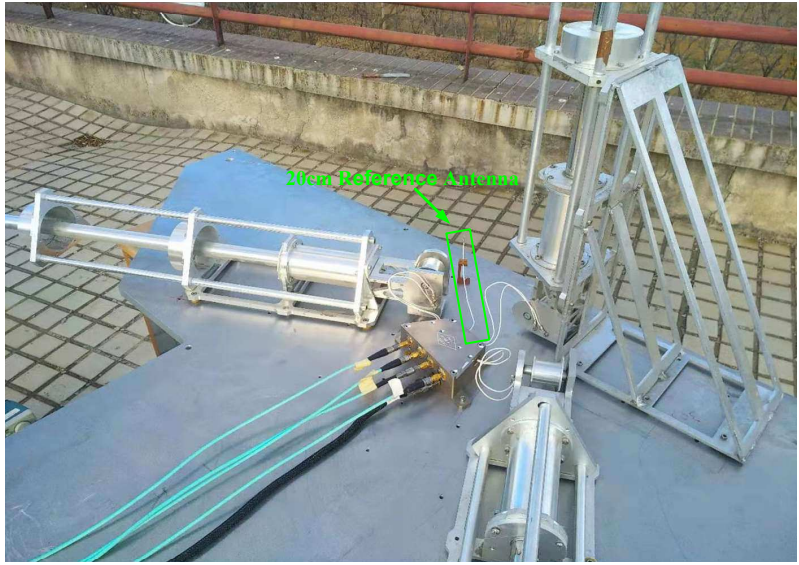}
   \caption{{\small Photograph of the LFRS antennas and the pre-amplifier.} }
   \label{fig:ms2019-0316fig2}
  \end{figure}

\subsection{Experimental Setup}

The engineering model which is form, fit and functionally the same as the flight model, was tested in the ground experiments. The experimental settings are shown in \autoref{fig:ms2019-0316fig2}. The EMI from the Chang'e-4 lunar lander was simulated by an arbitrary waveform generator which reproduced the lander noise recorded during the EMC test. Both the sinusoidal and wide-band signals generated by a vector signal generator were used to emulate the sky signals. The format of the emulated sky signals is listed in \autoref{Tab:Conf}.
\begin{table}[htbp]
  \centering
  \caption{ {\small Format of the emulated sky Signals.}}\label{Tab:Conf}
    \begin{tabular}{llllp{12.7em}}
    \toprule
    No.   & Signal Mode & Frequency & Amplitude  & \multicolumn{1}{l}{Remark} \\
    \midrule
    1     & \multirow{6}[2]{*}{Sinusoidal signals } & 1.8MHz  & 24dBm  & \multirow{6}[2]{*}{\tabincell{c}{Sinusoidal signals are generated\\ by arbitrary waveform generator} } \\
    2     &       & 5MHz  & 18dBm  & \multicolumn{1}{l}{} \\
    3     &       & 8MHz  & 18dBm  & \multicolumn{1}{l}{} \\
    4     &       & 10MHz  & 4dBm  & \multicolumn{1}{l}{} \\
    5     &       & 22.5MHz  & -8dBm  & \multicolumn{1}{l}{} \\
    6     &       & 38MHz  & -8dBm  & \multicolumn{1}{l}{} \\
    \midrule
    7     & \multirow{3}[2]{*}{\tabincell{c}{  \\  \\Wide-band signals}} & 1.75 MHz  & 20dBm  & \tabincell{c}{Bandwidth:100KHz\\generated by vector signal generator } \\
    8     &       & 5 MHz  & 20dBm  & \tabincell{c}{Bandwidth:1MHz\\generated by vector signal generator } \\
    9     &       & 20MHz  & 20dBm  & \tabincell{c}{Bandwidth:1MHz\\generated by vector signal generator } \\
    \bottomrule
    \end{tabular}%
  \label{tab:addlabel}%
\end{table}

\subsection{Experimental method}

 As can be seen in \autoref{Tab:Conf}, nine experiments were performed. Different rows in \autoref{Tab:Conf} mean different configuration of the emulated sky signal. Each configuration should be verified by an experiment. Each experiment was done in two steps. First the signals were recorded by the LFRS while the simulated noise was off. Next, the simulated noise was turned on, both the simulated noise and the emulated sky signal were recorded by the LFRS.

A method named adaptive interference cancellation is used to suppress the EMI from the lander(\citealt{Fridman+2001}). A separate, dedicated reference channel is designed in order to obtain an independent estimate of the EMI from lander. The 20 cm long antenna for the reference channel is installed very close to the lander, near the root of the 5 meter long antenna, as shown in \autoref{fig:ms2019-0316fig2}. The received signal from the short antenna is almost the EMI from the lander, because of the limited sensitivity for external signals. The signals received by the 5 meter long antenna, are be corrected using the signals received by the short reference antenna to suppress the EMI from the lander, the emulated sky signals with EMI and without EMI are compared to evaluate EMI mitigation.
For the data of every experiments, the first 50 groups of data are averaged as correction coefficient. And for every subsequent signal received by the antenna with a length of 5 meter is corrected using the corresponding signal received by the short reference antenna and the estimated correction coefficient. 75 groups of data(about 5 minute) are averaged as the result of the experiments.The above method we used to suppress the EMI from the lander is shown in Equation(\ref{equ:emi}).

\begin{equation}
{\begin{array}{*{20}{l}}

{C\mathop{{}}\nolimits_{{a}}{ \left( {f} \right) }=\frac{{{\mathop{ \sum }\limits_{{i=1}}^{{50}}{U\mathop{{}}\nolimits_{{i}}^{{a}}{ \left( {f} \right) }}}}}{{{\mathop{ \sum }\limits_{{i=1}}^{{50}}{U\mathop{{}}\nolimits_{{i}}^{{d}}{ \left( {f} \right) }}}}}}\\
{UC\mathop{{}}\nolimits_{{50+j}}^{{a}}{ \left( {f} \right) }=U\mathop{{}}\nolimits_{{50+j}}^{{a}}{ \left( {f} \right) }-
C\mathop{{}}\nolimits_{{a}}{ \left( {f} \right) } \times U\mathop{{}}\nolimits_{{50+j}}^{{d}}{ \left( {f} \right) }}\\
{UO\mathop{{}}\nolimits_{{}}^{{a}}{ \left( {f} \right) }=\frac{{1}}{{75}}{\mathop{ \sum }\limits_{{j=1}}^{{75}}{UC\mathop{{}}\nolimits_{{50+j}}^{{a}}}}{ \left( {f} \right) }}
\end{array}}
\label{equ:emi}
\end{equation}
where $U\mathop{{}}\nolimits_{{i}}^{{a}}{ \left( {f} \right) }$ is the $i$th group data output by LFRS which is a amplitude spectrum array of 5m long antenna $a$, $U\mathop{{}}\nolimits_{{i}}^{{d}}{ \left( {f} \right) }$ is the $i$th group data of reference antenna $d$, $ C\mathop{{}}\nolimits_{{a}}{ \left( {f} \right) }$  is the array of correction coefficient of antenna $a$, $UC\mathop{{}}\nolimits_{{50+j}}^{{a}}{ \left( {f} \right) }$ is the $50+j$th group of amplitude spectrum array which is corrected by reference antenna data and the array of correction coefficient,$UO\mathop{{}}\nolimits_{{}}^{{a}}{ \left( {f} \right) }$ represents the final amplitude spectrum array of 5m long antenna $a$.

The advantage of adaptive interference cancellation lies in keeping the structure of the signal-of-interest intact, the subtraction of an EMI estimate should not affect the wanted radio signal.
This kind of EMI cancellation is especially useful for the LFRS observations where the EMI and the signal-of-interest occupy the same frequency domain, but the effectiveness of the method depends also on the temporal stability of correction coefficient. Equation(\ref{equ:emi}) is workable under the assumption of constant correction coefficient during the later averaging. Furthermore, the effectiveness of this adaptive interference cancellation also depends on the sensitivity of the reference channel. There will be a loss of signal-of-interest after processing,if the auxiliary reference is sensitive enough to receive the external signals.

\subsection{Experimental procedure}

  \begin{figure}
   \centering
   \includegraphics[width=0.5\textwidth, angle=0]{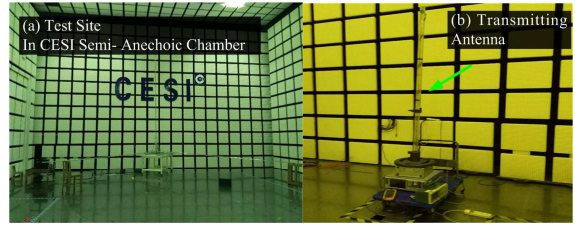}
   \caption{{\small Photographs of the test site CESI semi anechoic chamber.}}
   \label{fig:ms2019-0316fig3}
  \end{figure}
In order to prevent the low-frequency RFI from environment and simulate an open area test site, a semi-anechoic chamber located in CESI\footnote{ Abbreviation:CESI,China Electronic standardization Institute} was selected as the site of the ground experiments as showed in \autoref{fig:ms2019-0316fig3}(a)(b). The semi-anechoic chamber consists of a 23m(Length)$\times$ 14m(Width)$\times$ 9m(Height) shielded enclosure. The chamber is lined with hybrid absorbers(model number IP-130BLB) which consist of carbon loaded polystyrene foam absorber bonded to ferrite absorber backing. The guaranteed level of shielding effectiveness the chamber would provide over the frequency range of 100kHz to 18GHz is 100dB. Due to good electromagnetic shielding effectiveness of the semi-anechoic chamber, desired results of the ground experiments were achieved. The results and performance analysis are presented in the following section.

% reversion complete 2020.06.05

\section{ RESULTS AND PERFORMANCE ANALYSIS}
\subsection{RESULTS}
\begin{figure}[h]
  \centering
   \includegraphics[width=140mm]{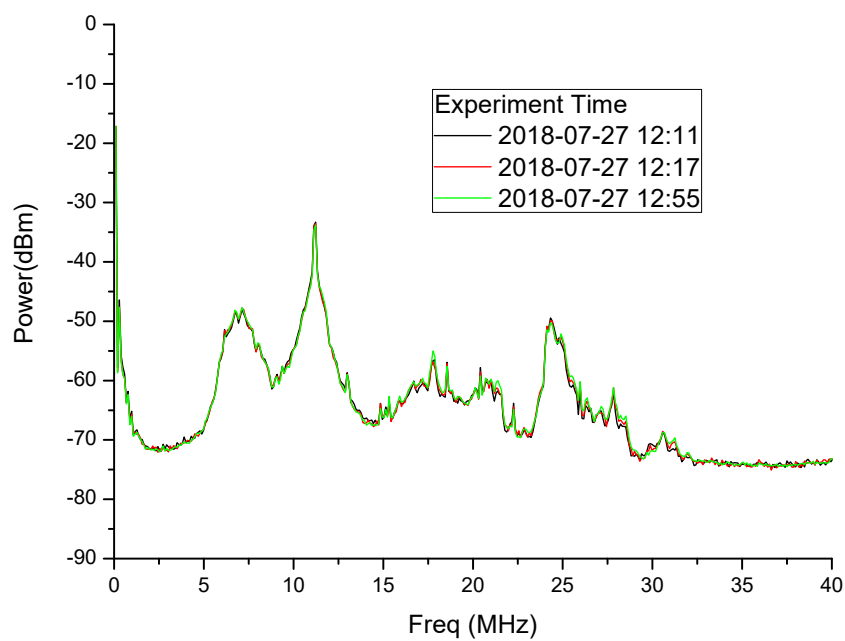}
  \caption{{\small Background signal spectra on site CESI semi-anechoic chamber.}}
     \label{fig:ms2019-0316fig4}
\end{figure}

At the beginning of the experiment, the background noise of the site was tested. \autoref{fig:ms2019-0316fig4} shows the background signal spectra received by the LFRS's 5 meter long antenna in the semi-anechoic chamber. The Y-axis represents the amplitude of the signal spectrum in dBm, while the X-axis shows the frequency of the signal spectrum in MHz.
Three tests were performed to check the characteristics of the background signals(black, red and green lines in \autoref{fig:ms2019-0316fig4}). As you can see by \autoref{fig:ms2019-0316fig4}, although there are still RFI from the computers, test instruments, power supply, etc., the power and frequency of the RFI does not appreciably change over observable time. For the above the reasons, the results from the experiments in the semi-anechoic chamber are used for analysis and performance evaluation of the LFRS.
% reversion complete 2020.06.28

 \begin{figure}
   \centering
   \includegraphics[width=1\textwidth]{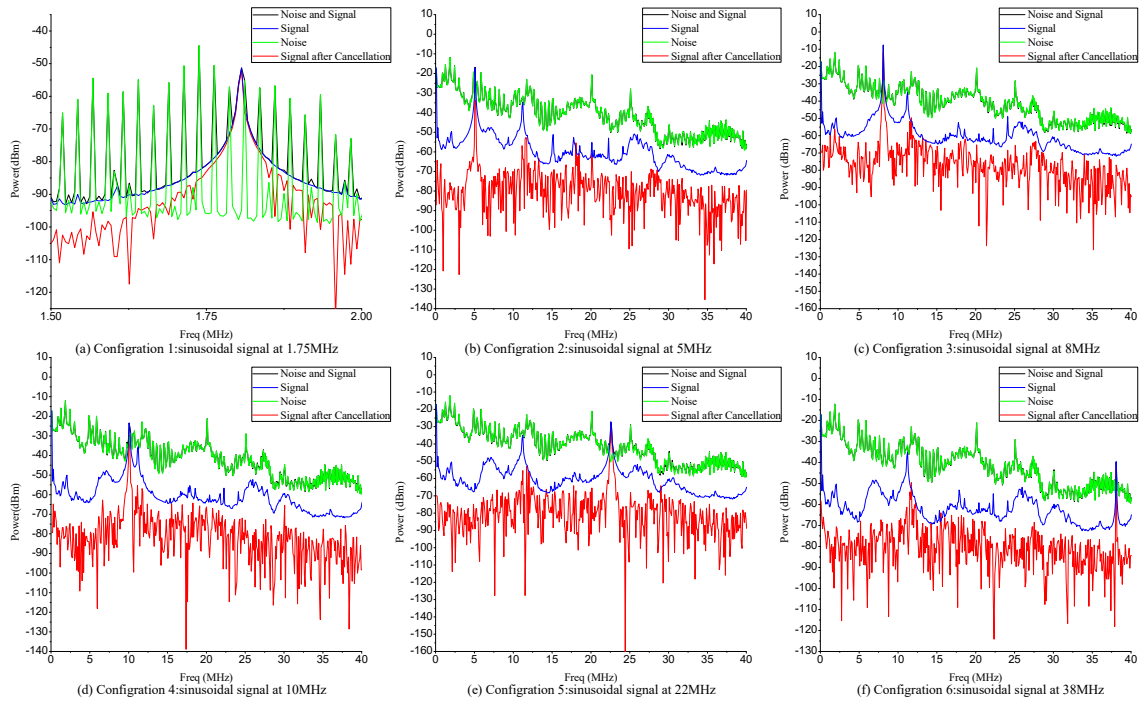}
   \caption{{\small Typical results of sinusoidal signals tests.}}
   \label{fig:ms2019-0316fig5}
  \end{figure}

 \autoref{fig:ms2019-0316fig5} shows the typical results of sinusoidal signals tests. According to the configuration of  \autoref{Tab:Conf}, six tests of sinusoidal signals mode were carried out.
The method which is briefly described in Section 3.2 was used for EMI mitigation. In \autoref{fig:ms2019-0316fig5} the blue lines indicate the spectrum of the generated signal described in \autoref{Tab:Conf}. The green lines show the spectra of lander noise recorded during the EMC test. The black lines represent mixed spectra of signal and lander noise. The red lines reflect the spectra after EMI mitigation. The position of the red line relative to the position of the black line could be used to see the effectiveness of EMI mitigation. As shown in  \autoref{fig:ms2019-0316fig5}, about 20 to 40dB noise reduction could be received  0.1-40MHz after EMI mitigation, while the sinusoidal signals are not severely distorted.

   \begin{figure}
   \centering
   \includegraphics[width=1\textwidth]{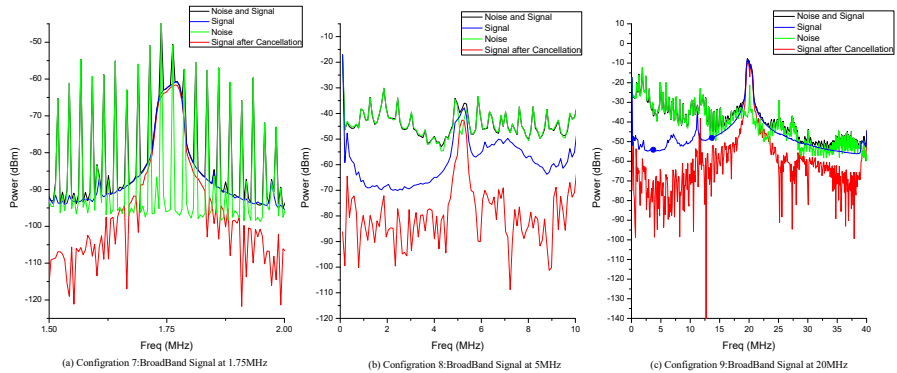}
   \caption{{\small Typical results of broadBand signals tests.} }
   \label{fig:ms2019-0316fig6}
  \end{figure}

 \autoref{fig:ms2019-0316fig6} show the typical results of broadband signal tests. As shown in\autoref{fig:ms2019-0316fig6}, about 20 to 40dB noise reduction could be received in 0.1-40MHz after EMI mitigation, while the broadband signals are not severely distorted.

\subsection{PERFORMANCE ANALYSIS}

   \begin{figure}
   \centering
   \includegraphics[width=0.8\textwidth]{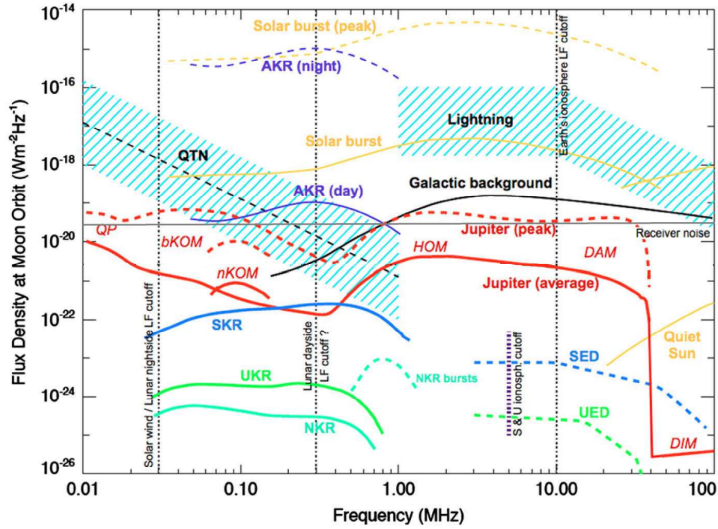}
   \caption{{\small Average and peak spectra of the various planetary magnetosphere and atmospheric radio components(\citealt{Zarka+2012}).} }
   \label{fig:ms2019-0316fig7}
  \end{figure}
Average and peak spectra of the various planetary magnetosphere and atmospheric radio components as they would be measured from the Moon surface or orbit are displayed in \autoref{fig:ms2019-0316fig7}. As stated in Section 1, the primary scientific objective of the LFRS is to probe radio emission from the solar burst on the surface of the Moon, so the performance analysis focus on the average and peak spectra of solar burst.

   \begin{figure}
   \centering
   \includegraphics[width=1\textwidth]{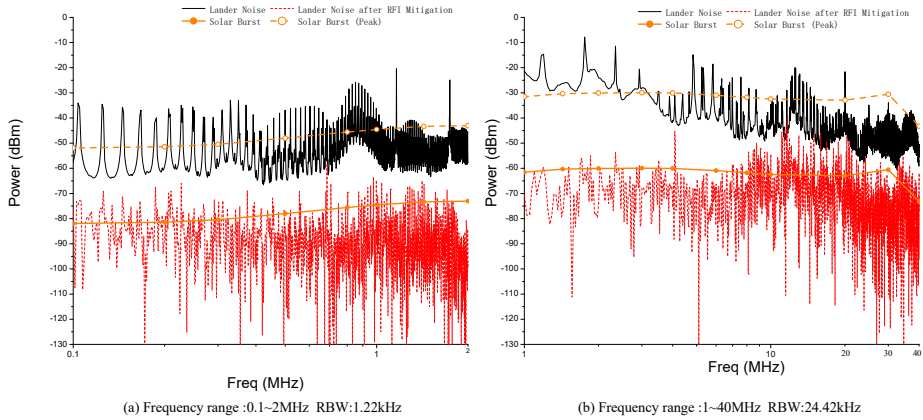}
   \caption{Power spectrum received by the LFRS on the Moon caused by solar burst and lander noise. }
   \label{fig:ms2019-0316fig8}
  \end{figure}

Because of the existence of lander noise, the detection capability of the LFRS is not only related to its technical performance, but also depends on the EMI mitigation method for the lander's noise suppression. In order to analyse the performance, the power of the signal received by the LFRS on the Moon caused by solar burst and lander noise are estimated and compared in
\autoref{fig:ms2019-0316fig8}.

 The power of the signal from solar burst (yellow lines in \autoref{fig:ms2019-0316fig8})could be estimated by solar burst flux density( solid or dashed yellow line in \autoref{fig:ms2019-0316fig7}) and the LFRS technical parameters using the equation (\ref{equ:sflux}), where $E$ is the electric field strength that produces voltage at the terminal of the antenna, $S$ is the flux density in $Wm\mathop{{}}\nolimits^{{-2}}Hz\mathop{{}}\nolimits^{{-1}}$ of the wave, $\eta$ is the intrinsic impedance of free space, $B$ is the bandwidth of the LFRS, $U\mathop{{}}\nolimits_{{a}}$ is the voltage at the terminal of the LFRS's antenna, $h\mathop{{}}\nolimits_{{e}}$ is the effective length of the antenna, $U\mathop{{}}\nolimits_{{preamp}}$ is the voltage outputted by pre-amplifier, $Z\mathop{{}}\nolimits_{{a}}$ is the impedance of the LFRS's antenna, $Z\mathop{{}}\nolimits_{{preamp}}$ is the impedance of the LFRS's pre-amplifier, $G\mathop{{}}\nolimits_{{preamp}}$ is the gain of the LFRS's pre-amplifier, $P\mathop{{}}\nolimits_{{out}}$ is the output power of the LFRS's receiver, $G\mathop{{}}\nolimits_{{rec}}$ is the gain of the LFRS's receiver, $L\mathop{{}}\nolimits_{{line}}$ is the insert loss of the cable between the pre-amplifier and the receiver, $Z\mathop{{}}\nolimits_{{L}}$ is the load resistance of the receiver.
\begin{equation}
{\begin{array}{*{20}{l}}
{E=\sqrt{{2 \eta BS}}}\\
{U\mathop{{}}\nolimits_{{a}}=E \times h\mathop{{}}\nolimits_{{e}}}\\
{U\mathop{{}}\nolimits_{{preamp}}=\frac{{U\mathop{{}}\nolimits_{{a}}Z\mathop{{}}\nolimits_{{preamp}}G\mathop{{}}\nolimits_{{preamp}}}}{{Z\mathop{{}}\nolimits_{{a}}+Z\mathop{{}}\nolimits_{{preamp}}}}}\\
{P\mathop{{}}\nolimits_{{out}}=\frac{{{ \left( {U\mathop{{}}\nolimits_{{preamp}}G\mathop{{}}\nolimits_{{rec}}L\mathop{{}}\nolimits_{{line}}} \right) }\mathop{{}}\nolimits^{{2}}}}{{Z\mathop{{}}\nolimits_{{L}}}}}\\
\end{array}}
\label{equ:sflux}
\end{equation}

 The power of the noise from the lander (black lines in \autoref{fig:ms2019-0316fig8})was recorded during the lander EMC test. The red lines in \autoref{fig:ms2019-0316fig8} represent the lander noise after EMI mitigation. The position of the red line relative to the position of the yellow line could be used to analyse detectability of solar burst. \autoref{fig:ms2019-0316fig8} shows that after EMI mitigation the solar burst with average flux density could be detected by the LFRS in some frequency range such as $1-6MHz$, while solar burst with peak flux density could be detected  by the LFRS in entire working frequency range of $0.1-40MHz$. In other words, when the radiation flux density of solar burst events or other similar events reach higher than $10\mathop{{}}\nolimits^{{-18}}Wm\mathop{{}}\nolimits^{{-2}}Hz\mathop{{}}\nolimits^{{-1}}$, these events might be detected by the LFRS.

\section{conclusions}

The properties, ground experiments and performance evaluation of the LFRS are briefly described in this paper. The results of the ground experiments show that the method using 20cm long antenna as reference antenna for EMI mitigation is helpful. The noise from the lander can by reduced by 20dB to 40dB, while the signals are not severely distorted in the frequency range of $0.1-40MHz$ with the above method. The detection capability and performance analysis show that after EMI mitigation solar burst with average flux density could be detected by the LFRS in some frequency range such as $1-6MHz$, while solar burst with peak flux density could be detected  by the LFRS in entire working frequency range of $0.1-40MHz$.
However, a lot of work needs to be done in the future lunar surface detection. Because of the difference between the  experimental environment and the real environment of the lunar surface, there may be a better way for EMI mitigation. Some suggestions to improve the EMI mitigation: after a period of observation on the lunar surface, the characteristics of the lander noise should be carefully studied. A more accurate model of the lander noise will be very helpful for the process and analysis of the spectra collected by the LFRS. Furthermore, more advanced and complicated algorithms can be tried to improve the detection capability of the LFRS(\cite{Fridman+2001}).

\begin{acknowledgements}
This work was funded by the Youth Innovation Promotion Association of CAS\footnote{ Abbreviation: CAS,Chinese Academy of Sciences} (Grant No. 2015334) and the National Natural Science Foundation of China(Grant No. 11941002).
The authors thank the Ground Application System of Lunar Exploration, National Astronomical Observatories and Institute of Electronics, CAS,  for their valuable and efficient assistance with providing the data and data calibration.

\end{acknowledgements}

\label{lastpage}

\end{document}